\documentclass{iopart}

\newcommand{\lie}{\mathcal{L}}

\begin{document}

\title[Pure indirect control of quantum system via accessor]
{Indirect control of quantum system via accessor: pure coherent
control without system excitation}

\author{H. C. Fu$^1$\footnote{E-mail: hcfu@szu.edu.cn},
Hui Dong$^2$, X. F. Liu$^3$,
C. P. Sun$^2$\footnote{E-mail: suncp@itp.ac.cn}}
\address{
$^1$ School of Physics Science and Technology, Shenzhen University, Shenzhen
518060, P.\,R.\,China \\
$^2$ Institute of Theoretical Physics, Chinese Academy of
Sciences, Beijing 100080, P.\,R.\,China\\
$^3$ Department of mathematics, Peking University, Beijing
100871, P.\,R.\,China}
\date{28/01/2008}

%\author{}
%\address{}
%\author{}
%\affiliation{Department of mathematics, Peking University, Beijing
%100871, P.\,R.\,China}
%\author{}
%\affiliation{Institute of Theoretical Physics, Chinese Academy of
%Sciences, Beijing 100080, P.\,R.\,China}
%\date{28/01/2008}

\begin{abstract}
A pure indirect control of quantum systems via quantum accessor is
investigated. In this control scheme, we do not apply any external
classical excitation fields on the controlled system and we control a
quantum system via a quantum accessor and classical control fields
control the accessor only. Complete controllability is investigated
for arbitrary finite dimensional quantum systems and exemplified
by 2 and 3 dimensional systems.
The scheme exhibits some advantages; it
uses less qubits in accessor and does not depend on the energy-level
structure of the controlled system.
\end{abstract}

%\submitto{\JPA}

\maketitle

\section{Introduction}

Quantum control is a coherence-preserving manipulation of a quantum
system, which enables a time evolution from an arbitrary initial  state to an
arbitrary target state [1-4]. It was first proposed by Huang {\em
et.\,al.} \cite{huang} in 1983 and was mainly used to control chemical
reaction in  its early days \cite{chem}. Recently
it has attracted much attention due to its connection to quantum
information processing. Actually the universality of quantum logic
gates can be understood from viewpoint of complete controllability in
quantum control \cite{gate}.
Conventional quantum control is the coherent control of quantum
systems using classical external fields. Controllability of this
{\em semi-classical} control is well studied \cite{con1}, especially the
complete controllability of finite dimensional quantum systems using
Lie algebra method \cite{fu1,fu2} graph method \cite{tur} and
transfer graph method \cite{ran}.
Lie algebra approach plays important role in the investigation
in both the classical control \cite{1972} and the quantum control.

In some circumstances in quantum information processing, there is
need to control qubits using quantum controllers such as quantum
accessor and environment. For example, in connection with the
fundamental limit of quantum information processing and influence of
decoherence to quantum control, we have proposed an indirect scheme
for quantum control where the controller is also quantum \cite{xue}.
To avoid switching the couplings between qubits, Zhou {\it et. al.}
introduced the so-called encoded qubits to realize the universal
quantum computation with local manipulation of physical qubits only
\cite{zhou1}. Here the physical qubits do not involve the quantum
computation and play the role of quantum controllers. Recently
Hodges {\em et. al.} proposed an universal indirect control of
nuclear spins using a single electron spin acting as an accessor
driven by microwave irradiation of resolved anisotropic hyperfine
\cite{cqed}, which has important application for spin based solid
state quantum information processing. Therefore the control of
quantum systems using quantum controllers has significant
application in quantum information processing  and has attracted
much attention recently. Authors of this paper proposed the
conception of the {\em indirect control} of quantum systems where
the quantum systems are controlled via a quantum accessor and the
classical control fields control the accessor only \cite{indirect1}.
Similar works were proposed in different context \cite{rrdd} for
spin-1/2 particles. Romano \cite{rom2} and Pechen \cite{env}
considered the incoherent control induced by environment modeled as
quantum radiation fields.

In our previous paper \cite{indirect1}, we proposed a scheme
for the control of an arbitrary finite dimensional
quantum system using a quantum accessor modeled as a
qubit chain with XY-type neighborhood coupling. We find the
conditions of way of coupling between the controlled system and
accessor and the minimal length of qubit chain to ensure the complete control
of the controlled system. However, besides the classical control
fields controlling on the accessor, we also apply a constant classical
field on the controlled system to excite the system through dipole
interaction. Without the excitation field, the system is not completely
controllable and for the 2-dimensional case, underlying
Lie algebra is the Symplectic algebra sp(4),
rather than su(4). Another disadvantage of
this scheme is that the controllability depends on the structure of energy
levels of the controlled system. In
this paper we shall remove the excitation field and propose a {\em pure}
indirect control scheme where the external control fields control the
accessor {\em only}. We shall see that, in comparison with the scheme in
\cite{indirect1}, the new scheme proposed in this paper exhibits
some advantages besides the removal of excitation field, for
example, it uses less qubits of the accessor for complete control of
the controlled system and there is no particular requirements on the
structure of the energy levels of controlled system.

The remaining part of this paper is organized as follows: we
formulate the control system without system excitation in Sec.\,2
and then introduce the {\em selection} operators and apply it to the
study of controllability of two energy level system in Sec.\,3. The case of
3-dimensional system is investigated in Sec.\,4. The general
approach of controllability of indirect control of arbitrary
finite-dimensional systems is investigated in Sec.\,5. We conclude
in Sec.\,6.

\section{Indirect control system}

In this section we shall formulate the indirect control system and
fix the notations we will use later on. Suppose that the system to be
controlled is an $N$-dimensional quantum system described by the
following Hamiltonian
\begin{equation}
H_S = \sum_{i=1}^{N}E_i e_{ii} = \sum_{i=1}^{N-1}\epsilon_i h_i,
\end{equation}
where $E_i$'s are eigen energy of the system, $e_{ij}$ is an
$N\times N$ matrix with matrix elements
$(e_{ij})_{kl}=\delta_{ik}\delta_{jl}$, $h_i = e_{ii}-e_{i+1,i+1}$
 are Cartan generators of the Lie algebra su($N$)
 and $\epsilon_i \equiv E_1+E_2+\cdots+E_{i}$. Here we have assumed that
$\mbox{tr} H_S=0$ without losing generality.

Note that we do not apply
any external classical excitation field on the system S as we did in
\cite{indirect1}. The excitation field, although it is a constant field,
makes the indirect control in \cite{indirect1} not really pure indirect.

The quantum accessor is modeled as a qubit chain with XY-type
neighborhood coupling
\begin{eqnarray}
&&H_A = H_A^0 + H_A^I, \nonumber \\
&&H_A^0 = \sum_{i=0}^M \hbar \omega_i \sigma_z^i, \ \ \ \ \ H_A^I =
\sum_{i=1}^{M-1}c_i\sigma_x^i\sigma_{x}^{i+1},
\end{eqnarray}
where $c_i\neq 0$ and
\begin{equation}
   \sigma_\alpha^i = 1\otimes \cdots \otimes 1\otimes \sigma_\alpha \otimes
   1\otimes \cdots \otimes 1,
\end{equation}
namely the $\sigma_\alpha$ on the site $i$ and 1 on any other
sites.

The system and the accessor are coupled as
\begin{equation}
H_I = \sum_{\{\alpha_i\}}\left[\sum_{j=1}^{N-1}\sum_{k=0,\pm 1}
g_{(\alpha_i)}^{j(k)} s_j^{k}\right]\otimes
\sigma^1_{\alpha_1}\cdots\sigma^M_{\alpha_M}, \label{interhamil}
\end{equation}
where $\{\alpha_i\}=\{\alpha_1,\alpha_2,\cdots,\alpha_M\}$ and each
$\alpha_i=x,y,z$ rather than just $x, y$ as in previous paper \cite{indirect1},
$s_j^k$ is defined as
\begin{equation}
s_j^k =\left\{
\begin{array}{ll}
x_j   &    \mbox{ when } k=1; \\
h_j   &    \mbox{ when } k=0; \\
y_j   &    \mbox{ when } k=-1,
\end{array}
\right.
\end{equation}
and
\begin{eqnarray}
&&  x_j = e_{j,j+1}+e_{j+1,j}, \nonumber \\
&&  y_j = i\left(e_{j,j+1}-e_{j+1,j}\right),
\end{eqnarray}
along with $h_j$ constitute the Chevalley basis of the Lie algebra
su($N$) \cite{lie}.

It is known that when we remove the excitation field, the indirect system
is not completely controllable if the $\alpha_i = x, y$ only \cite{indirect1}. In the
case of indirect control of 2-level system, the Lie algebra is sp(4) with
dimension 10, rather than the su(4) \cite{indirect1}. However, as the example we presented in
\cite{indirect1}, we can rotate the system to remove the excitation
field, but as price paid the interaction Hamiltonian includes $\sigma_z$
for the accessor part. So this is why we includes $\alpha_i = z$ in
the coupling Hamiltonian (\ref{interhamil}).

We suppose that we can control each qubit of the accessor completely
through external classical fields. The complete control of each
qubit in a qubit chain can be achieved via global manipulation
\cite{zhou2,sonia}. Therefore the total control system is
\begin{eqnarray}
&& H=H_0+\sum_{j=1}^{M}
\left[f_j(t)\sigma_x^j+f^{\prime}_j(t)\sigma_y^j\right], \nonumber \\
&& H_0=H_S+H_A+H_I, \label{controlsystem}
\end{eqnarray}
where $f_j(t)$ and $f^{\prime}_j(t)$ are two independent classical
control fields.

In the  rest of this paper we shall investigate the complete
controllability of the indirect control scheme (\ref{controlsystem}),
namely in what conditions the system is completely
controllable.
More precisely, in what conditions the Lie algebra generated by
the skew-Hermitian operators
$iH_0$, $i1\otimes\sigma_x^j$ and  $i1\otimes\sigma_y^j$ $(j=1,2,\cdots,M)$
is su$(2^M N)$, or its dimension is $\left(2^M N\right)^2-1$.

\section{Selection operators and Indirect control of single qubit}

To prove the complete controllability, we define the so-called {\em
selection operators} $S^k_{xy}$ and $S^k_{yx}$ acting on the Pauli's
operators of $k$-th qubit of the accessor
\begin{equation}
S^k_{xy}=\frac{1}{4}\mbox{ad}_{i\sigma_x^k}\mbox{ad}_{i\sigma_y^k}, \ \ \ \
S^k_{yx}=\frac{1}{4}\mbox{ad}_{i\sigma_y^k}\mbox{ad}_{i\sigma_x^k}
\end{equation}
where $\mbox{ad}_{i\sigma_x^k}$ is the adjoint representation of
the Lie algebra su(2) of the $k$-th qubit
\begin{equation}
\mbox{ad}_{X}(Y)=[X,Y], \ \ \forall X,Y \in \mbox{su(2)}.
\label{adj}
\end{equation}
From definition (\ref{adj}), it follows
\begin{eqnarray}
&& S_{xy}^k(*)=\frac{1}{4}\left[ i\sigma _x^k,\left[ i\sigma
_y^k,*\right] \right] ,
\nonumber \\
&& S_{yx}^k(*)=\frac{1}{4}\left[ i\sigma
_y^k,\left[ i\sigma _x^k,*\right] \right].
\end{eqnarray}
It is easy to prove that
\begin{eqnarray}
S_{xy}^k(i\sigma_\alpha^k) &=&\left\{
\begin{array}{ll}
i\sigma _y^k, & \alpha=x; \\
0, & \alpha=y,z;
\end{array}
\right. \\
S_{yx}^k(i\sigma_\alpha^k) &=&\left\{
\begin{array}{ll}
i\sigma _x^k, & \alpha=y; \\
0, & \alpha=x,z;
\end{array}
\right.
\end{eqnarray}
namely, $S_{xy}^k$ transforms the $i\sigma^k_x$ to $i\sigma^k_y$,
$S_{yx}^k$ transforms the $i\sigma^k_y$ to $i\sigma^k_x$ and
they annihilate any others. Or in other words, $S_{xy}^{k}$ can {\em
select} the $\sigma_x^k$ and change it to $\sigma_y^k$ from any
linear combination of Pauli's matrices.

Now we show how to use those operators in the investigation of
complete controllability with the 2-dimensional system as an
example. Here both the system and accessor are single qubit. The
Hamiltonian of the system and accessor is as follows
\begin{eqnarray}
H_0 &=&\hbar \omega _S\sigma _z\otimes 1+\hbar\omega _A(1\otimes \sigma _z)+ \nonumber\\
&&+\left(g_{xx}\sigma _x + g_{yx}\sigma_y +g_{zx}\sigma_z \right)\otimes
\sigma_x  \nonumber \\
&&+\left(g_{xy}\sigma_x+g_{yy}\sigma_y+g_{zy}\sigma_z\right)\otimes \sigma_y
\nonumber \\
&& +\left(g_{xz}\sigma _x+g_{yz}\sigma_y+g_{zz}\sigma_{z}\right)\otimes
\sigma_z.
\end{eqnarray}
We suppose we can control the accessor fully
\begin{equation}
H_C=f_1(t)1\otimes \sigma _x+f_2(t)1\otimes \sigma _y,
\end{equation}
where $f_1(t)$ and $f_2(t)$ are two independent classical
control fields to control the accessor. The Lie algebra
generators are $iH_0$, $i1\otimes \sigma_x$ and $i1\otimes \sigma_y$
and the generated Lie algebra is denoted by ${\cal L}$. It is
obvious that
\begin{equation}
-2^{-1}\left[i1\otimes \sigma_x, i1\otimes \sigma_y\right] =
i1\otimes\sigma_z \in {\cal L}.
\end{equation}
So we can subtract the second term in $H_0$ and obtain the Lie
algebra element $iH'_0 \equiv iH_0-i \hbar \omega _A(1\otimes \sigma
_z)\in {\cal L}$.

Now we apply the selection operators on the element $iH'_0$,
yielding
\begin{eqnarray}
S_{xy}(iH'_0) &=& i\left( g_{xx}\sigma _x+g_{yx}\sigma
_y+g_{zx}\sigma
_z\right) \otimes \sigma_y \in \mathcal{L},  \label{7} \\
S_{yx}(iH'_0)&=&i\left( g_{xy}\sigma _x+g_{yy}\sigma _y+g_{zy}\sigma
_z\right) \otimes \sigma_x\in \mathcal{L}.  \label{4}
\end{eqnarray}
In fact, by evaluating the commutation relation of (\ref{7},\ref{4})
with the generators $\sigma_x$ and $\sigma_y$ of accessor in (\ref{7})
and (\ref{4}) can be changed to any $\sigma_\alpha$ ($\alpha=x,y,z$).

We further subtract the terms (\ref{7}) and (\ref{4}) from $iH^\prime_0$ and
then calculate its commutation relation with $i1\otimes\sigma_y$. We find
\begin{equation}
i\left( g_{xz}\sigma _x+g_{yz}\sigma _y+g_{zz}\sigma _z\right)
\otimes \sigma _\alpha \in \mathcal{L}.  \label{6}
\end{equation}
If the following condition
\begin{equation}
\det \left(
\begin{array}{ccc}
g_{xx} & g_{yx} & g_{zx} \\
g_{xy} & g_{yy} & g_{zy} \\
g_{xz} & g_{yz} & g_{zz}
\end{array}
\right) \neq 0  \label{2dcondition}
\end{equation}
is satisfied, we find nine Lie algebra elements
\begin{equation}
i\sigma _\alpha \otimes \sigma _\beta \in \mathcal{L},
\end{equation}
where $\alpha ,\beta =x,y,z$. The condition (\ref{2dcondition}) can
be achieved by choosing, for example, $g_{xx}=g_{yy}=g_{zz}=1$ and any others zero.
As we already have $%
i1\otimes \sigma _\alpha \in \mathcal{L}$, so we only need to prove $i\sigma
_\alpha \otimes 1\in \mathcal{L}$. For this purpose, we evaluate
\begin{eqnarray}
&&-2^{-1}\left[ i\sigma _x\otimes \sigma _x,i\sigma _y\otimes \sigma
_x\right]
=i\sigma _z\otimes 1\in \mathcal{L}, \\
&&2^{-1}\left[ i\sigma _x\otimes \sigma _x,i\sigma _z\otimes \sigma
_x\right]
=i\sigma _y\otimes 1\in \mathcal{L}, \\
&&2^{-1}\left[ i\sigma _y\otimes 1,i\sigma _z\otimes 1\right]
=i\sigma _x\otimes 1\in \mathcal{L}.
\end{eqnarray}

In summary, the generated Lie algebra has fifteen generators
$i\sigma_\alpha\otimes\sigma_\beta$ where $\alpha,\beta=x,y,z,0$ (
$\sigma_0 \equiv 1$) and $\alpha,\beta$ cannot be 0 simultaneously,
and they generate the Lie algebra su($4$). Therefore the single
qubit system is completely controllable under the condition
(\ref{2dcondition}).

\section{Control of 3-dimensional system}

In this section we turn to the indirect control of 3-dimensional quantum
system. The Hamiltonian takes the following form
\begin{eqnarray*}
H_0 &=&H_S+H_A+H_{SA} \\
H_S &=&\sum_{i=1}^3E_ie_{ii}=E_1h_1+\left( E_1+E_2\right) h_2 \\
H_A &=&\hbar \omega _1\sigma _z^1+\hbar \omega _2\sigma _z^2+c\sigma
_x^1\sigma _x^2 \\
H_{SA} &=&\sum_{\alpha ,\beta =x,y,z}\left( g_{\alpha \beta
}^{1(0)}h_1+g_{\alpha \beta }^{2(0)}h_2+g_{\alpha \beta
}^{1(1)}x_1+g_{\alpha \beta }^{2(1)}x_2\right. \\
&&\left. +g_{\alpha \beta }^{1(-1)}y_1+g_{\alpha \beta }^{2(-1)}y_2\right)
\otimes \sigma _\alpha ^1\sigma _\beta ^2
\end{eqnarray*}
where $h_1=e_{11}-e_{22}$ and $h_2=e_{22}-e_{33}$ are Cartan
elements of Lie algebra su(3), and $x_i=e_{i,i+1}+e_{i+1,i}$ and
$y_i=i\left( e_{i,i+1}-e_{i+1,i}\right) $ ($i=1, 2$)
are Chevelley basis of
su(3) corresponding
positive and negative simple roots, respectively.
%In the complex coefficients $%
%g_{\alpha \beta }^{1(1)}$, the first upper index represents index labeled
%the Chevelley basis and the second one in bracket represents type of
%Chevelley basis
%\[
%k=\left\{
%\begin{array}{ll}
%-1 & \text{for }y \\
%0 & \text{for }h \\
%1 & \text{for }x
%\end{array}
%\right.
%\]
The complete control system is
\[
H=H_0+\sum_{k=1}^2 \left(f_k(t)1\otimes \sigma_\alpha ^k+ f_k^\prime
(t)1\otimes\sigma_\alpha^k\right),
\]
where $f_k(t)$ and $f_k^\prime(t)$ are classical control fields.

It is easy to see that $i1\otimes\sigma_z^k\in {\cal L}$. So we can
subtract the free Hamiltonian of the accessor form $H_0$ and obtain
the following Lie algebra element
\begin{equation}
H_0^\prime = H_0 - \left(\hbar \omega _1\sigma _z^1+\hbar \omega
_2\sigma _z^2\right)\in {\cal L}.
\end{equation}
%Its commutation relations with $i\left( 1\otimes \sigma _\alpha
%^k\right) $.
It is easy to check that
\begin{eqnarray}
S_{yx}^2 S_{yx}^1 \left( iH_0^\prime\right) &=& i\left(
g_{yy}^{1(0)}h_1+g_{yy}^{2(0)}h_2+g_{yy}^{1(1)}x_1\right.  \nonumber  \\
&&
\left.+g_{yy}^{2(1)}x_2+g_{yy}^{1(-1)}y_1+g_{yy}^{2(-1)}y_2\right)
\otimes \sigma _x^1\sigma _x^2 \in \lie, \label{31}
\\
\lefteqn{S_{xy}^2S_{yx}^1\left( iH_0^\prime\right) = i\left(
g_{yx}^{1(0)}h_1+g_{yx}^{2(0)}h_2+g_{yx}^{1(1)}x_1+\right. }
\nonumber
\\
&& \left.g_{yx}^{2(1)}x_2+g_{yx}^{1(-1)}y_1+g_{yx}^{2(-1)}y_2\right)
\otimes \sigma _x^1\sigma _y^2 \in \lie,\label{32}
\\
\lefteqn{\left( 1-S_{xy}^2-S_{yx}^2\right) S_{yx}^1\left(
iH_0^\prime\right) = i\left( g_{yz}^{1(0)}h_1 \right.} \nonumber
\\
&&
\left.+g_{yz}^{2(0)}h_2+g_{yz}^{1(1)}x_1+g_{yz}^{2(1)}x_2+g_{yz}^{1(-1)}y_1+\right.
\nonumber \\
&& \left.g_{yz}^{2(-1)}y_2\right) \otimes \sigma _x^1\sigma _z^2 \in
\lie. \label{33}
\end{eqnarray}

After proper commutation with the external interaction Hamiltonian,
we can change the accessor part in Eqs.(\ref{31}-\ref{33}) to
$\sigma^1_{y}\sigma^2_y$,
$\sigma^1_{y}\sigma^2_x$ and $\sigma^1_{y}\sigma^2_z$, respectively.
Then we
subtract those Lie algebra elements from $iH_0^\prime$
and obtain the following Lie algebra element
\begin{eqnarray}
iH_0^{\prime\prime} &=& c\sigma _x^1\sigma
_x^2  +\sum_{\alpha =x,z}\sum_{\beta =x,y,z}\left( g_{\alpha \beta
}^{1(0)}h_1+g_{\alpha \beta
}^{2(0)}h_2+g_{\alpha \beta }^{1(1)}x_1 \right.\nonumber \\
& & \left.+ g_{\alpha \beta }^{2(1)}x_2+g_{\alpha \beta
}^{1(-1)}y_1+g_{\alpha \beta }^{2(-1)}y_2\right) \otimes \sigma
_\alpha ^1\sigma _\beta ^2.
\end{eqnarray}
To remove the term $c\sigma_x^1 \sigma_x^2$ from
$iH_0^{\prime\prime}$, we evaluate the commutation relation between
$iH_0^{\prime\prime}$ and $i\left( 1\otimes \sigma _x^1\right) $,
yielding
\begin{eqnarray}
iH_0^{\prime\prime\prime}&=&-2^{-1}\left[iH_0^{\prime\prime},
i1\otimes \sigma_x^1\right] \nonumber \\
&=& i\sum_{\beta=x,y,z}\left( g_{z\beta }^{1(0)}h_1+g_{z\beta
}^{2(0)}h_2+g_{z\beta }^{1(1)}x_1+g_{z\beta
}^{2(1)}x_2 \right. \nonumber \\
& & +\left. g_{z\beta }^{1(-1)}y_1+g_{z\beta }^{2(-1)}y_2\right)
\otimes \sigma _y^1\sigma _\beta ^2\in \mathcal{L}.
\end{eqnarray}
Then we can use the same trick as in (\ref{31}-\ref{33}) to prove
\begin{eqnarray}
S_{yx}^2\left( iH^{\prime\prime\prime }\right)  &=& i\left(
g_{zy}^{1(0)}h_1+g_{zy}^{2(0)}h_2+ g_{zy}^{1(1)}x_1 +g_{zy}^{2(1)}x_2\right.
 \nonumber \\
&& \left.  + g_{zy}^{1(-1)}y_1+g_{zy}^{2(-1)}y_2\right) \otimes
\sigma_x^1\sigma _x^2\in {\cal L},  \label{34}
\\
%%%%%%%%%%%%%%%%%%%%%%%%%%%%%%%%%%%%%%%%%%%%%%%%%%%%%%%%%%%%
\lefteqn{S_{xy}^2\left( iH^{\prime\prime\prime }\right)=i\left(
g_{zx}^{1(0)}h_1+g_{zx}^{2(0)}h_2+ g_{zx}^{1(1)}x_1 +
g_{zx}^{2(1)}x_2\right. } \nonumber \\
&& \left.  + g_{zx}^{1(-1)}y_1+g_{zx}^{2(-1)}y_2\right)
\otimes \sigma_x^1\sigma_y^2\in \cal{L}, \label{35}  \\
%%%%%%%%%%%%%%%%%%%%%%%%%%%%%%%%%%%%%%%%%%%%%%%%%%%%%%%%%%%%%%%
\lefteqn{\left( 1-S_{xy}^2-S_{yx}^2\right) \left(
iH^{\prime\prime\prime}\right)
=i\left(g_{zz}^{1(0)}h_1+g_{zz}^{2(0)}h_2+ \right.} \nonumber\\
&& \left.g_{zz}^{1(1)}x_1
+g_{zz}^{2(1)}x_2+g_{zz}^{1(-1)}y_1+g_{zz}^{2(-1)}y_2\right)\nonumber\\
&& \otimes \sigma _x^1\sigma _z^2 \in {\cal L}. \label{36}
\end{eqnarray}

Now we have found six independent Lie algebra elements
Eqs.(\ref{31}-\ref{33}) and Eqs.(\ref{34}-\ref{36}), in which
the accessor part can be changed to the same $\sigma^1_\alpha
\sigma^2_\beta$ ($\alpha,\beta=x,y,z$) by
evaluating proper commutation with the
external interaction Hamiltonian. If the
coefficients satisfy the following condition
\begin{equation}
\det \left(
\begin{array}{llllll}
g_{yx}^{1(0)} & g_{yx}^{2(0)} & g_{yx}^{1(1)} & g_{yx}^{2(1)} &
g_{yx}^{1(-1)} & g_{yx}^{2(-1)} \\
g_{yy}^{1(0)} & g_{yy}^{2(0)} & g_{yy}^{1(1)} & g_{yy}^{2(1)} &
g_{yy}^{1(-1)} & g_{yy}^{2(-1)} \\
g_{yz}^{1(0)} & g_{yz}^{2(0)} & g_{yz}^{1(1)} & g_{yz}^{2(1)} &
g_{yz}^{1(-1)} & g_{yz}^{2(-1)} \\
g_{zx}^{1(0)} & g_{zx}^{2(0)} & g_{zx}^{1(1)} & g_{zx}^{2(1)} &
g_{zx}^{1(-1)} & g_{zx}^{2(-1)} \\
g_{zy}^{1(0)} & g_{zy}^{2(0)} & g_{zy}^{1(1)} & g_{zy}^{2(1)} &
g_{zy}^{1(-1)} & g_{zy}^{2(-1)} \\
g_{zz}^{1(0)} & g_{zz}^{2(0)} & g_{zz}^{1(1)} & g_{zz}^{2(1)} &
g_{zz}^{1(-1)} & g_{zz}^{2(-1)}
\end{array}
\right) \neq 0,
\end{equation}
we have that all the elements
\begin{equation}
h_k\otimes \sigma _\alpha ^1\sigma _\beta ^2\in \mathcal{L}, \ \ \ \
x_k\otimes \sigma _\alpha ^1\sigma _\beta ^2\in \mathcal{L}, \ \ \ \
y_k\otimes \sigma _\alpha ^1\sigma _\beta ^2\in \mathcal{L},
\end{equation}
where $k=1,2$ and $\alpha,\beta=x,y,z$.

 So we need further to prove
$1\otimes \sigma _\alpha ^1\sigma _\beta ^2\in
\mathcal{L}$ and $h_k\otimes 1\in \mathcal{L}$, $x_k\otimes 1\in \mathcal{L}$%
, $y_k\otimes 1\in \mathcal{L}$. To this end let us evaluate
\begin{eqnarray}
-\frac 12\left[ ih_k\otimes \sigma _\alpha ^1\sigma _\beta ^2,ix_k\otimes
\sigma _\alpha ^1\sigma _\beta ^2\right]  &=&y_k\otimes 1\in \mathcal{L}, \\
\frac 12\left[ ih_k\otimes \sigma _\alpha ^1\sigma _\beta ^2,y_k\otimes
\sigma _\alpha ^1\sigma _\beta ^2\right]  &=&ix_k\otimes 1\in \mathcal{L}, \\
\frac 12\left[ ix_k\otimes \sigma _\alpha ^1\sigma _\beta
^2,y_k\otimes \sigma _\alpha ^1\sigma _\beta ^2\right]
&=&ih_k\otimes 1\in \mathcal{L}.
\end{eqnarray}

In this 3-dimensional system case, we choose all coefficients
$g^{j(k)}_{x\alpha}=0$. Then from $iH_0^{\prime\prime}$ we subtract
the Lie algebra elements (\ref{31}-\ref{33}, \ref{34}-\ref{36})
with the same accessor part $\sigma^1_\alpha
\sigma^2_\beta$ ($\alpha,\beta=x,y,z$) and find
\begin{equation}
i 1\otimes \sigma_x^1\sigma_x^2 \in \lie
\end{equation}
from which we have $i1\otimes \sigma_\alpha^1 \sigma_\beta^2 \in
\lie$. So we have proved the complete controllability of
3-dimensional quantum systems.

Here we would like to note that the
Lie algebra su(3) has 6 Chevalley basis and therefore we need 6 equations
to decouple the terms in Hamiltonian. However, there are nine elements of type
$i\sigma_\alpha^1 \sigma_\beta^2$.

%%%%%%%%%%%%%%%%%%%%%%%%%%%%%%%%%%%%%%%%%%%%%%%%%%%%%%%%%%%%%%%%%%%%%%%%%
\section{Complete controllability of finite dimensional quantum system}
%%%%%%%%%%%%%%%%%%%%%%%%%%%%%%%%%%%%%%%%%%%%%%%%%%%%%%%%%%%%%%%%%%%%%%%%%

With experience built in previous two sections, we shall
generally investigate the complete
controllability of arbitrary finite dimensional quantum systems
in this section. In
the interaction Hamiltonian $H_I$ there are $3^M$ coupling terms
\begin{equation}
\left[\sum_{j=1}^{N-1}\sum_{k} g_{\{\alpha_i\}}^{j(k)}
s_j^{k}\right] \otimes \sigma^1_{{\alpha}_1}
\sigma^2_{{\alpha}_2}\cdots \sigma^M_{{\alpha}_M}. \label{nomial}
\end{equation}
Here we call them {\em nomial} for convenience. Notice that the
accessor part in each nomial is labeled by an index set
$\{\alpha_i\}$. In the forthcoming part of this paper we use symbol
$ \{\alpha_i | n \}$ to denote this index set, in which the number
of $\sigma_z$ in the nomials is not less than $n$. It is obvious for
$ \{\alpha_i | n \}$, there are
\begin{equation}
\left(\begin{array}{l} M \\ n
\end{array}\right) 2^{M-n}
\end{equation}
nomials in which there are $n$ $\sigma_z$'s. One can easily check
that the sum of those numbers gives rise to $3^M$ using binomial
formula, the total coupling terms in $H_I$, as we expected.

In the following we shall first prove each nomials (\ref{nomial}) is
in Lie algebra ${\cal L}$ and then prove the generated Lie algebra
is su($N 2^M$).

\subsection{Decoupling $iH_I$ to nomials}

We first prove that each terms in $H_I$ is in the Lie algebra ${\cal
L}$. We shall prove this recursively according to the number of
$\sigma_z$ in each nomial.

We first notice that the element $iH_A^0 \in {\cal L}$, so the
element $iH^{(0)} \equiv iH_0-iH_A^0 \in \lie$. Without losing
generality, we suppose $M \ge 3$ hereafter.

As the first step, we would like to {\em select} the terms without
$\sigma_z$ in the qubit chain. For this purpose, we first annihilate
$iH_A^I$ and the nomials with $\sigma_z$'s in $iH_I$ from $iH^{0}$
by evaluating the commutation relations
\begin{eqnarray}
iH^{(0)1} &\equiv&  \left[i\sigma_z^M, \left[i\sigma_z^{M-1},
\cdots, \left[ i\sigma^1_z,
iH^{(0)}\right]\cdots\right]\right] \nonumber \\
&=&i2^M\sum_{[\![\alpha_i]\!]}(-1)^{\Delta_{[\![\alpha_i]\!]}}\left[\sum_{j=1}^{N-1}\sum_{k}
g_{[\![\alpha_i]\!]}^{j(k)} s_j^{k}\right] \otimes \nonumber \\
&& \sigma^1_{\overline{\alpha}_1}
\sigma^2_{\overline{\alpha}_2}\cdots \sigma^M_{\overline{\alpha}_M}
\in \lie, \label{xyelements0}
\end{eqnarray}
where we have used the symbol $[\![\alpha_i]\!]$ to denote the index
set with each $\alpha_i = x,y$ only, and $\Delta_{[\![\alpha_i]\!]}$
is the number of $x$ in $[\![\alpha_i]\!]$. Note that the terms
$iH_A^I$ is also annihilated as each terms in it has only two
neighborhood qubits.

As each index in (\ref{xyelements0}) is either $x$ or $y$, we can
use selection operators to pick up each nomial in
Eq.(\ref{xyelements0})
\begin{eqnarray}
\lefteqn{ S^{M}_{\beta_{M},\overline{\beta}_{M}}
S^{M-1}_{\beta_{M-1},\overline{\beta}_{M-1}} \cdots
S^{1}_{\beta_1,\overline{\beta}_1} \left(iH^{(0)1}\right) = }\nonumber \\
&& i\left[\sum_{j=1}^{N-1}\sum_{k} g_{\{\beta_i\}}^{j(k)}
s_j^{k}\right] \otimes  \sigma^1_{\overline{\beta}_1}
\sigma^2_{\overline{\beta}_2}\cdots \sigma^M_{\overline{\beta}_M}
\in \lie,    \label{xyelements}
\end{eqnarray}
where $\beta_i, \overline{\beta}_i = x, y$ and
\begin{equation}
\overline{\beta}_i=\left\{ \begin{array}{ll} x, & \mbox{if }\beta_i=y;\\
y, & \mbox{if }\beta_i=x.
\end{array}
\right. \label{xyelements1}
\end{equation}
So the Eq.(\ref{xyelements}) implies that we have $2^M$ Lie algebra elements
in which the $\beta_i$ is either $x$ or $y$.

As the second step, we further deprive nomials with just one
$\sigma_z$ in qubit chain. We first evaluate the proper commutation
with external interaction Hamiltonian to change the $\sigma^k_{\bar{\beta}_k}$
to $\sigma^k_{\beta_k}$ in Eq.(\ref{xyelements}) and then subtract them
from $iH^{(0)}$. We obtain the following Lie
algebra element
\begin{eqnarray}
iH^{(1)} &\equiv& iH_S+iH_A^I  \nonumber \\
&& +i\sum_{\{\alpha_i|1\}}\left[\sum_{j=1}^{N-1}\sum_{k}
g_{\{\alpha_i|1\} }^{j(k)} s_j^{k}\right] \otimes
\sigma^1_{\alpha_1} \sigma^2_{\alpha_2}\cdots \sigma^M_{\alpha_M}
 \in \lie. \label{41}
\end{eqnarray}
Without losing generality, we consider the case where $\alpha_1=z$.
As in step 1, we would like to annihilate the term $iH_A^I$ and all
nomials that has $\sigma_z$ in sites other than site 1. For this
purpose, we evaluate commutation relation on the site $1$ with
$i\sigma^1_x$ and other sites with $\sigma^j_z$. Those $M$
operations change $\sigma^1_z$ to $\sigma^1_y$ and other
$\sigma^n_x$ to $\sigma^n_y$ or vise versa for $2\leq n\leq M$. We
have
\begin{eqnarray}
iH^{(1)1} &\equiv& \left[i\sigma_z^M, \cdots, \left[i\sigma_z^{2},
\left[ i\sigma^1_x,
iH^{(1)}\right]\cdots\right]\right] \nonumber \\
&=& i 2^M \sum_{\{\alpha_i|1\} \atop \alpha_1=z}
(-1)^{\Delta_{\{\alpha_i|1\}}}\left[\sum_{j=1}^{N-1}\sum_{k}
g_{\{\alpha_i|1\} }^{j(k)}
s_j^{k}\right] \otimes \nonumber \\
&&\sigma^1_{y} \sigma^2_{\overline{\alpha}_2}\cdots
\sigma^M_{\overline{\alpha}_M} \in \lie.
\end{eqnarray}
where $\alpha_i=x$ or $y$ for $i=2,3,\cdots,M$. As each site of the
accessor is either $\sigma_x$ or $\sigma_y$, we can use selection
operators to find $2^{M-1}$ elements of ${\cal L}$
\begin{equation}
 i\left[\sum_{j=1}^{N-1}\sum_{k} g_{\{ z,\beta_2,
 \cdots,\beta_M|1\}}^{j(k)} s_j^{k}\right] \otimes \sigma^1_{y}
\sigma^2_{\beta_2}\cdots \sigma^M_{\beta_M} \in \lie. \label{one-z}
\end{equation}

In fact, we can use the same method to prove that nomials with only
one $\alpha_k=z$ on the site $k$ are elements of Lie algebra
$\cal{L}$. In total we have $M2^{M-1}$ such type Lie algebra
elements.

We can now subtract the element
\begin{equation}
 i\left[\sum_{j=1}^{N-1}\sum_{k} g_{\{ z,\beta_2,
 \cdots,\beta_M|1\}}^{j(k)} s_j^{k}\right] \otimes \sigma^1_{z}
\sigma^2_{\beta_2}\cdots \sigma^M_{\beta_M} \in \lie
\end{equation}
which can be obtained from the commutation relation of $\sigma_x^1$
with (\ref{one-z}), and obtain an
element of ${\cal L}$ which takes the same form of (\ref{41}) but
there are at least two $z$ in $[\![ \alpha_i ]\!]$
\begin{eqnarray}
\lefteqn{iH^{(2)} \equiv iH_S+iH_A^I +} \nonumber \\
&& i\sum_{\{\alpha_i|2\}}\left[\sum_{j=1}^{N-1}\sum_{k}
g_{\{\alpha_i|2\} }^{j(k)} s_j^{k}\right] \otimes
\sigma^1_{\alpha_1} \sigma^2_{\alpha_2}\cdots \sigma^M_{\alpha_M}
\nonumber \\
 && \in \lie.
\end{eqnarray}
Suppose that $\alpha_m=\alpha_n=z$ ($m\neq n$). Then we can evaluate
the commutation relation of $iH^{(2)}$ with $i\sigma_x^m$,
$i\sigma_x^n$ and $i\sigma_z^k$ ($k\neq m,n$), one can easily prove
that the element with two $z$ in the Lie algebra ${\cal L}$.

Following the procedure recursively on the number of $z$ in $\{
\alpha_i \}$, we can prove all the elements
\begin{equation}
i \left[\sum_{j=1}^{N-1}\sum_{k} g_{\{\alpha_i \}}^{j(k)}
s_j^{k}\right] \otimes \sigma^1_{\alpha_1} \sigma^2_{\alpha_2}\cdots
\sigma^M_{\alpha_M} \in \lie, \label{threeelement}
\end{equation}
where each $\alpha_i = x,y,z$. There are $3^M$ such elements.

As each nomial of the type (\ref{threeelement}) is a linear combination
of $3(N-1)$ elements $x_i$, $y_i$ and $h_i$ ($i=1,2,\cdots, N-1$), so
we require that the number of qubits is big enough such that
\begin{equation}
3^M \geq 3(N-1),
\label{qnumber}
\end{equation}
and then choose $3(N-1)$ elements of type (\ref{threeelement}).
Then we further require the determinant of the coefficient matrix
\begin{equation} \label{condition2}
\mbox{det}\left(g^{i(k)}_{\{\alpha_i\}} \right) \neq 0,
\end{equation}
we find that the elements
\begin{eqnarray}
&&ix_i\otimes \sigma^1_{\alpha_1} \sigma^2_{\alpha_2}\cdots
\sigma^M_{\alpha_M}, \\
&&iy_i\otimes \sigma^1_{\alpha_1} \sigma^2_{\alpha_2}\cdots
\sigma^M_{\alpha_M}, \\
&& ih_i\otimes \sigma^1_{\alpha_1} \sigma^2_{\alpha_2}\cdots
\sigma^M_{\alpha_M},
\end{eqnarray}
are Lie algebra elements. Namely, all nomials in interaction
Hamiltonian $iH_I$ are decoupled and each term is in the Lie algebra
${\cal L}$.

\subsection{System operators as Lie algebra elements}

It is easy to see that
\begin{eqnarray}
&&\frac{1}{2}\left[ ix_i\otimes \sigma^1_{\alpha_1} \sigma^2_{\alpha_2}\cdots
\sigma^M_{\alpha_M}, y_i\otimes \sigma^1_{\alpha_1}
\sigma^2_{\alpha_2}\cdots
\sigma^M_{\alpha_M}\right]  = i h_i\otimes 1_A \in {\cal L}, \\
&& -\frac{1}{2}\left[ ih_i\otimes \sigma^1_{\alpha_1} \sigma^2_{\alpha_2}\cdots
\sigma^M_{\alpha_M},  ix_i\otimes \sigma^1_{\alpha_1}
\sigma^2_{\alpha_2}\cdots
\sigma^M_{\alpha_M}\right]= y_i\otimes 1_A \in {\cal L},
\\
&& -\frac{1}{2}\left[ ih_i\otimes \sigma^1_{\alpha_1} \sigma^2_{\alpha_2}\cdots
\sigma^M_{\alpha_M},  iy_i\otimes \sigma^1_{\alpha_1}
\sigma^2_{\alpha_2}\cdots
\sigma^M_{\alpha_M}\right]=i y_i\otimes 1_A \in {\cal L}.
\end{eqnarray}
From those Chevalley basis elements corresponding to simple roots of Lie
algebra su($N$), we can further construct the standard Cartan basis
of su($N$) corresponding any other positive and negative roots. We
have in total $N^2-1$ such basis elements of $\lie$.

\subsection{Accessor elements}

Above discussions mean that the Hamiltonian $iH_S^0$ is an element
of ${\cal L}$. So subtracting this element along with $iH_I$ and
$iH_A^0$, we find that $iH_A^I \in {\cal L}$.

It is easy to see that
\begin{equation}
\left[ \left[ iH_A^I,
i1\otimes\sigma_y^1\right],i1\otimes\sigma_y^1\right]=-i4c_1(1_S\otimes\sigma_x^1\sigma_x^2)\in
\lie
\end{equation}
thanks to the condition $c_1\neq 0$. We further have that
\begin{eqnarray}
\lefteqn{\left[\left[ iH_A^I-i c_1 1_S\otimes\sigma_x^1\sigma_x^2,
i1\otimes\sigma_y^2\right],i1\otimes\sigma_y^2\right]} \nonumber
\\
&&=-i4c_2(1_S\otimes\sigma_x^2\sigma_x^3)\in \lie
\end{eqnarray}
since $c_2\neq 0$. Repeating this process we can prove that
\begin{equation}
i\left(1_S\otimes\sigma_x^j\sigma_x^{j+1}\right)\in \lie, \ \ \ \
j=1,2,\cdots,M-1.
\end{equation}
Then form the Lemma 2 in Ref.\cite{indirect1}, we find that
\begin{equation}
i(1_S\otimes\sigma_{[\alpha]}) \in \lie, \ \ \ \ [\alpha]\neq
(0,0,\cdots,0)
\end{equation}
The number of those type of elements is $4^M-1$.

\subsection{Complete controllability}

So far we have proved that if the conditions (\ref{qnumber}) and
(\ref{condition2}) are
satisfied, the following elements are in Lie algebra $\lie$
\begin{eqnarray}
&& h_i\otimes 1_A \in \lie, \ \ \ \
   ix_i\otimes 1_A \in \lie, \ \ \ \
   iy_i\otimes 1_A \in \lie, \nonumber \\
&& h_i\otimes \sigma_{[\alpha]} \in \lie,\ \ \ \
   ix_i\otimes \sigma_{[\alpha]} \in \lie,\ \ \ \
   iy_i\otimes \sigma_{[\alpha]} \in \lie, \nonumber \\
&& i(1_S\otimes\sigma_{[\alpha]}) \in \lie,
\end{eqnarray}
and their corresponding Cartan basis elements.
The total number of those Lie algebra elements is
\begin{equation}
(N^2-1)+(N^2-1)(4^M-1)+(4^M-1) = (2^MN)^2-1,
\end{equation}
which is the dimension of Lie algebra su($2^MN$). This proves the
complete controllability of the indirect control system
(\ref{controlsystem}).

\section{Conclusion}

In this paper we have proposed a scheme for the indirect control of
finite dimensional quantum systems via quantum accessor modeled as
a qubit chain with XY-type coupling. The main results of this paper
are as follows:
\begin{itemize}
\item Different from our previous
paper \cite{indirect1}, we do not need to apply an excitation classical
field on the controlled system. So this scheme is a {\em pure} indirect
control in the sense that the classical control fields control the
accessor only.

\item  The minimal number $M$ for the
complete control of the controlled system is determined by condition
(\ref{qnumber}), while in previous scheme \cite{indirect1}, the
minimal $M$ is determined by $2^M \geq 2(N-1)$.
It is obvious that the scheme proposed here requires less qubits in
accessor in comparison to the proposal \cite{indirect1}.

\item
We also notice that in the process of
decoupling the interaction Hamiltonian (see Sec.5.1) we do not put any
requirements on the structure of energy-level of the
controlled system, while in \cite{indirect1}, the indirect controllability
reduces to the semi-classical control investigated in \cite{fu1,fu2} which
depends on the energy-level structure of the controlled system.

\item From Eq.(\ref{condition2}) we find that the controllability
is determined by the way of coupling of controlled system and accessor.
So in a practical control protocol we can design a simplest coupling of the controlled system and
accessor to ensure the complete control of the controlled system, according
to the condition (\ref{condition2}).
\end{itemize}

Therefore we believe the scheme in this paper has wider
applicability. As further works we would like to study the
concrete control protocol of the indirect control, and examine the graph
connectivity for assessing the controllability of quantum systems,
as well as applications in quantum information processing.

\section*{Acknowledgement}

This work is supported by the NSFC by grants No.\,10675058,
No.\,90233018, No.\,10474144 and No.\,60433050, and the NFRPC by
grants No.\,2006CB921205 and N0.\,2005CB724508.

\end{document}